\documentclass[11pt]{article}
\usepackage{epsfig}
\usepackage{latexsym}
\begin{document}
\thispagestyle{empty}
\begin{flushright}
hep-lat/0112024
\end{flushright}
\begin{center}
\vspace*{18mm}
{\LARGE Setting the scale for the 
L\"uscher-Weisz action} 
\vskip15mm
{\bf Christof Gattringer${\,}^\dagger$, Roland Hoffmann \\
and Stefan Schaefer}
\vskip4mm
Institut f\"ur Theoretische Physik \\
Universit\"at Regensburg \\
93040 Regensburg, Germany
\vskip25mm
\begin{abstract}
We study the quark-antiquark potential of quenched SU(3) lattice
gauge theory with the L\"uscher-Weisz action. After blocking the gauge 
fields with the recently proposed
hypercubic transformation we compute the Sommer
parameter, extract the lattice spacing $a$ and set the scale 
at 6 different values of the gauge coupling in a range from
$a = 0.084$ fm to 0.136 fm.
\end{abstract}
\vskip5mm
{\sl To appear in Physical Review D.}
\vskip5mm
\end{center}
\vskip10mm 
\nopagebreak \begin{flushleft} \rule{2 in}{0.03cm}
\\ {\footnotesize \ 
${}^\dagger$ Supported by the Austrian Academy of Sciences (APART 654).}
\end{flushleft}
\newpage
\setcounter{page}{1}

Recently a noticeable revival of the interest in improved
gauge actions took place. It was observed that improved gauge actions
make the numerical problems less severe when implementing chiral fermions.
The underlying mechanism for the numerical improvement is a suppression 
of ultraviolet fluctuations of the gauge field. 
An example is the use of the Iwasaki and other improved gauge actions
for domain wall fermions \cite{Or01} and also 
using the perfect gauge action \cite{perfgauge}
is an integral part of constructing the fixed point Dirac operator
\cite{fixpd}. Recently a systematic expansion of a solution of the
Ginsparg-Wilson equation \cite{GiWi}, the so-called chirally
improved fermion, was proposed and implemented \cite{Gaetal}. Also 
there it was found that using the improved gauge action is numerically 
advantageous for the Dirac operator.
In particular the L\"uscher-Weisz action \cite{LuWe} with coefficients
from tadpole improved perturbation theory \cite{Aletal95,LeMa93} was used.
Subsequently the instanton content of the QCD vacuum was studied for the 
L\"uscher-Weisz action in \cite{Gaetal2}. 

In this letter we report on our results for the static potential and the
lattice scale for the L\"uscher-Weisz action in order to make the use of
this action easily accessible to the community. We furthermore test the 
recently proposed method of hypercubic blocking \cite{hypblock} which
was found \cite{hypblock,hyppotential} to
improve the statistical accuracy in the determination of the static
potential by an order of magnitude. We analyze quenched
ensembles at 6 different values of the gauge coupling and compute the 
Sommer parameter \cite{sommer1,sommer} and the lattice spacing $a$. 
This results 
in a precise determination of the lattice scale in a range between 
$a = 0.084$ fm to $a = 0.136$ fm and using a fit to
our data even beyond this interval. Together with our results for the 
secondary couplings $\beta_2$ and $\beta_3$ 
this letter provides all ingredients 
necessary for using the L\"uscher-Weisz action at typical lattice spacings  
of state of the art simulations. 

In addition to the plaquette term of the Wilson gauge action, the 
L\"uscher-Weisz action includes a sum over all $2 \! \times \! 1$
rectangles and a sum over all parallelograms, i.e.~all possible 
closed loops of length 6 along the edges of all 3-cubes. 
Explicitly the action reads 
\begin{eqnarray}
S[U] & = & \beta_1 \sum_{pl} \frac{1}{3} \; \mbox{Re~Tr} \; [ 1 - U_{pl} ] 
\; + \; 
\beta_2 \sum_{rt} \frac{1}{3} \; \mbox{Re~Tr} \; [ 1 - U_{rt} ] 
\nonumber 
\\
& + & 
\beta_3 \sum_{pg} \frac{1}{3} \; \mbox{Re~Tr} \; [ 1 - U_{pg} ] \; ,
\label{sgauge}
\end{eqnarray}
where $\beta_1$ is the principal parameter while  $\beta_2$ and $\beta_3$ can
be computed \cite{Aletal95}
from $\beta_1$  using one loop perturbation theory and tadpole improvement  
\cite{LeMa93}, 
\begin{equation}
\beta_2 \; = \; - \; \frac{ \beta_1}{ 20 \; u_0^2} \; 
[ 1 + 0.4805 \, \alpha ]
\; \; , \; \;
\beta_3 \; = \; - \; \frac{ \beta_1}{u_0^2} \; 0.03325 \, \alpha
\; ,
\end{equation}
with
\begin{equation}
u_0 \; = \; \Big( \frac{1}{3} \mbox{Re~Tr} \langle U_{pl} \rangle 
\Big)^{1/4} \; \; , \; \; \alpha \; = \; - \;
\frac{ \ln \Big(\frac{1}{3} \mbox{Re~Tr} \langle U_{pl} \rangle 
\Big)}{3.06839} \; .
\end{equation}
The couplings $\beta_2, \beta_3$ are 
determined self-consistently from $u_0$  and
$\alpha$ for a given $\beta_1$. In Table~\ref{rundat} we list 
the values of the $\beta_i$ used for our ensembles and our results 
for the expectation value of the plaquette 
${u_0}^4 = \mbox{Re~Tr} 
\langle U_{pl} \rangle/3$. The sample size at each value of 
$\beta_1$ is 200 configurations on $16^4$ lattices. The update was 
done with a mix of Metropolis and over-relaxation sweeps. 

\begin{table}[h]
\begin{center}
\hspace*{-2mm}
\begin{tabular}{c|cccccc}
\small
$\beta_1$ & 8.00 & 8.10 & 8.20 & 8.30 & 8.45 & 8.60 \\
\hline
${u_0}^4$ 
&$\!0.62107(3)\!\!$&$\!0.62894(3)\!\!$&$\!0.63599(3)\!\!$&$\!0.64252(3)\!\!$&
$\!0.65176(3)\!\!$&$\!0.66018(3)\!\!$\\
$\beta_2$ 
& $-0.54574$ & $-0.54745$ & $-0.54998$ & $-0.55332$ & $-0.55773$ & $-0.56345$\\
$\beta_3$ 
& $-0.05252$ & $-0.05120$ & $-0.05020$ & $-0.04953$ & $-0.04829$ & $-0.04755$
\normalsize
\end{tabular}
\end{center}
\caption{Parameters for the L\"uscher-Weisz action.
We list the values of the $\beta_i$ and the 
expectation value of the plaquette ${u_0}^4 = \mbox{Re~Tr}
\langle U_{pl} \rangle/3$.
\label{rundat}}
\end{table}

Before measuring the potential we applied the hypercubic blocking 
transformation proposed in \cite{hypblock}. Hypercubic blocking mixes only
gauge links from the hypercubes attached to the target link 
and has less impact on the short distance
properties of the gauge fields than previously used smearing methods. The 
hypercubic blocking transformation proceeds in three steps \cite{hypblock}:
\begin{eqnarray}
&\!\!\!\!\!\!\!\!\!\!& \overline{V}_{i,\mu ; \nu \, \rho } 
 \; = \; 
{\cal P }_{SU(3)} 
\left[ ( 1-\alpha _{3} ) U_{i,\mu } \; + \; \frac{\alpha _{3}}{2}
\sum _{\pm \eta \neq \rho ,\nu ,\mu }
U_{i,\eta }U_{i+\hat{\eta },\mu }U_{i+\hat{\mu },\eta }^{\dagger }
\right] \; , 
\nonumber 
\\
&\!\!\!\!\!\!\!\!\!\!& \widetilde{V}_{i,\mu ;\nu }  
 \; = \; 
{\cal P }_{SU(3)}
\left[ ( 1-\alpha _{2} ) U_{i,\mu } \; + \; \frac{\alpha _{2}}{4}
\sum _{\pm \rho \neq \nu ,\mu }
\overline{V}_{i,\rho ; \nu \, \mu } 
\overline{V}_{i+\hat{\rho },\mu ;\rho \, \nu }
\overline{V}_{i+\hat{\mu },\rho ;\nu \, \mu }^{\,\dagger }
\right] \; ,
\nonumber 
\\ 
&\!\!\!\!\!\!\!\!\!\!& V_{i,\mu }
 \; = \; 
{\cal P }_{SU(3)}
\left[ ( 1-\alpha _{1} ) U_{i,\mu }+\frac{\alpha _{1}}{6}
\sum _{\pm \nu \neq \mu }
\widetilde{V}_{i,\nu ;\mu } \widetilde{V}_{i+\hat{\nu },\mu ;\nu }
\widetilde{V}_{i+\hat{\mu },\nu ;\mu }^{\dagger }
\right]\, .
\label{blocking}
\end{eqnarray}
In the first step intermediate fields $\overline{V}_{i,\mu ; \nu \, \rho }$
are created from the thin-link variables $U_{i,\mu}$ (indices $i$ run over 
all sites of the lattice and $\mu,\nu,\rho$ and $\eta$ 
over the four directions).
In the second step the intermediate fields  
$\overline{V}_{i,\mu ; \nu \, \rho }$ are blocked into a second set of 
intermediate fields $\widetilde{V}_{i,\mu ;\nu }$
which in the third step are transformed into the 
final fields $V_{i,\mu }$. The restrictions on the indices 
$\mu,\nu$ and $\rho$ implemented in the sums in Eqs.~(\ref{blocking}) 
ensure that 
$V_{i,\mu }$ contains only contributions from the hypercubes attached to
the link $(i,\mu)$. By ${ \cal P}_{SU(3)}$ we denote the projection of
the sums back to elements of SU(3). The parameters $\alpha_1, \alpha_2$ and 
$\alpha_3$ determine the admixture of staples in each step of the
blocking process. These parameters 
were optimized \cite{hypblock} to minimize the
fluctuations of the plaquette. Their values are given by 
$\alpha_1 = 0.75, \alpha_2 = 0.6$ and $\alpha_3 = 0.3$. 

We measured the static potential on the smeared configurations using planar
Wilson loops $W(r,t)$ of size $r \times t$ with both $t$ and $r$ 
ranging from 1 to 10. We fitted the expectation values of the Wilson loops
to a sum of two exponentials 
$c_1 \exp( -V(r) t ) \, + \, c_2 \exp( - E^\prime t)$ 
in a range of $t = 2,3 \, ... \, 9$. 
The second exponential  
takes into account the contribution from excited states 
$E^\prime > V(r)$ and from 
the first term we directly obtain the potential $V(r)$ for two static sources
at distance $r$. As a cross check we also computed for some of the ensembles
the potential for the raw, unblocked configurations. We find that the results
are compatible within error bars but the statistical fluctuations, 
in particular at larger values of $r$ and $t$, are much more severe for the 
raw configurations. 
\begin{figure}[h]
\begin{center}
\vspace{3mm}
\hspace*{-8mm}
\epsfig{file=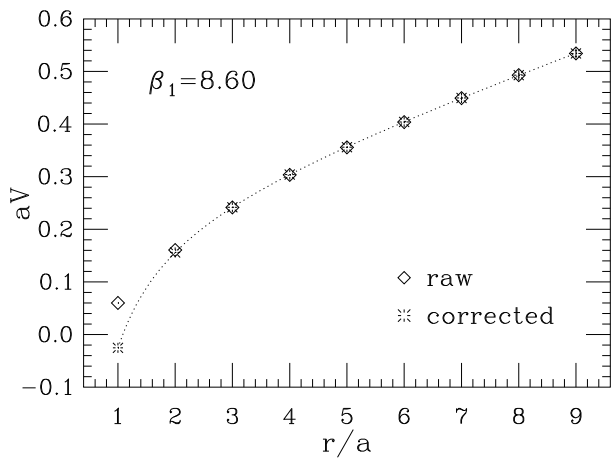,width=8cm,clip}
\caption{The static potential for the $\beta_1 = 8.60$
ensemble. The diamonds represent the raw data from the two parameter fit on
the Wilson loops and the bursts are the values corrected for 
the short distance effect of the hypercubic blocking. The full curve
is the parameterization (\ref{contpot}).
\label{potential860}}
\end{center}
\end{figure}

In Fig.~\ref{potential860} we show our results for the 
static potential for the $\beta_1 = 8.60$ ensemble. The smooth curve is the 
standard infrared parameterization for the continuum potential,
\begin{equation}
V(r) \; = \; C \; - \; A/r \; + \; \sigma r \; ,
\label{contpot}
\end{equation}
with constants $C,A$ and $\sigma$.
The diamonds are the values for the potential 
obtained from the Wilson loops. The error bars are smaller than the
symbols. For small distances,
one finds a noticeable deviation from the Coulomb behavior $- A/r$. This 
deviation is an effect of the hypercubic blocking. However, this effect can 
be computed perturbatively and the obtained deviation from the Coulomb
potential is used to introduce a fourth fit parameter in the potential fit
\cite{hyppotential}. Subtracting this perturbative part gives the 
corrected data which we represent by bursts. 
From the parameters $A$ and $\sigma$ we computed the Sommer parameter $r_0$ 
\cite{sommer} in lattice units $a$ and assuming $r_0 = 0.5$ fm we extracted 
the lattice spacing $a$. The result is very stable under variation of the
fit range for the potential. Even the $r/a=1$ measurement can be included.
The final result is the weighted average over all fit ranges $[ar_{min},ar_{max}]$
with $r_{min}\in\{1,2,3\}$ and $r_{max}\in\{7,8,9\}$.
We give the results for the lattice spacing and 
the Sommer parameter in Table \ref{results}.

\vspace*{4mm}

\begin{table}[h]
\begin{center}
\hspace*{-2mm}
\begin{tabular}{c|cccccc}
$\beta_1$ & 8.00 & 8.10 & 8.20 & 8.30 & 8.45 & 8.60 \\
\hline
$r_0/a$ 
& $\!3.688(37)\!$ & $\!4.015(34)\!$ & $\!4.362(41)\!$ & 
$\!4.741(49)\!$ & $\!5.289(66)\!$ & $\!5.967(70)\!$ \\
$a$ [fm] 
& 0.136(1) & 0.125(1) & 0.115(1) & 0.105(1) & 0.095(1) & 0.084(1) 
\end{tabular}
\end{center}
\vspace*{-2mm}
\caption{Results for the Sommer parameter $r_0$ in lattice units and 
the corresponding values for the lattice spacing $a$ when $r_0$ is assumed 
as $0.5 \rm{fm}$.
\label{results}}
\end{table}

\vspace*{4mm}

Without the improved fit $r_0$ can still
be determined using (\ref{contpot}). In this case, the lower limit of the fit
must be chosen so that the region where the HYP smearing distorts the potential
is excluded, as can bee seen from Figure \ref{potential860}, i.e. $r_{min}\in\{2,3\}$. The results of
an analysis based on the three parameter fit
(\ref{contpot}) are in perfect agreement with those given in the above table
(for e.g. $\beta_1=8.30$ one obtains $r_0=4.732(43)$). Thus
the only benefit from the improved fit is that no data points have to be
excluded. This shows that the effect of the HYP smearing on the short-distance
static quark potential
is under good perturbative control.

\pagebreak

In order to make the Sommer parameter and the lattice spacing available also
for other values of $\beta_1$ we fit our data to a functional form
based on the $\beta$-function as proposed in \cite{sommer}. We find
\begin{equation}
\ln(r_0/a) \; = \; 1.55354 \; + \; 0.79840 \, (\beta_1 - 8.3) \; - \;
0.09533 \, (\beta_1 - 8.3)^2 \; .
\label{fitfunction}
\end{equation} 

\begin{figure}[!h]
\begin{center}
\epsfig{file=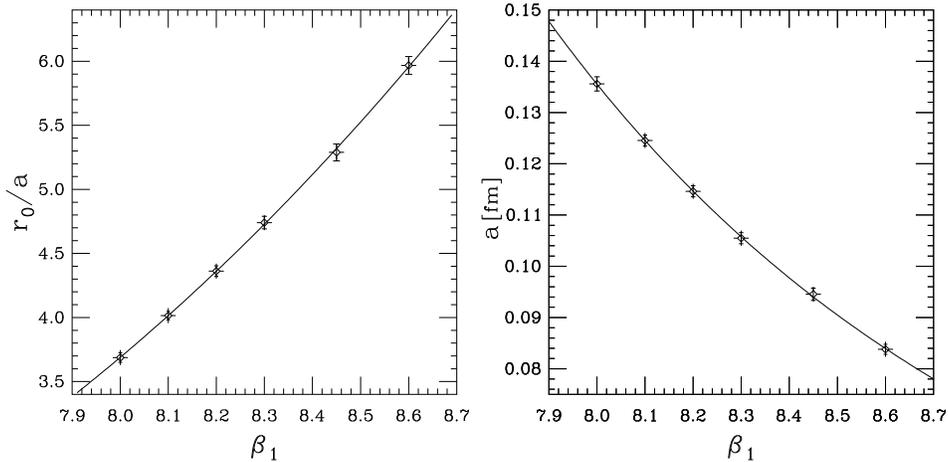,width=12.5cm,clip}
\caption{Results for the Sommer parameter in lattice units 
(left-hand side plot) and the lattice spacing in fermi 
(right-hand side plot) as a function of $\beta_1$. The full curves are 
the interpolations of the data with the function (\ref{fitfunction}).
\label{r0aplot}}
\end{center}
\end{figure}

In Fig.~\ref{r0aplot} we compare our numerical data for $r_0/a$ and $a$ 
(again assuming $r_0 = 0.5$ fm) to the curve (\ref{fitfunction}). It is 
obvious that the data are well described by our parameterization. Furthermore 
when extending the plot range to values of $\beta_1$ as small as 
$\beta_1 = 6.8$ we find that our results are in good agreement with 
the data computed for very coarse lattices in \cite{Aletal95}, 
i.e.~$a = 0.24$ fm at $\beta_1 = 7.4$, $a = 0.33$ fm at $\beta_1 = 7.1$
and  $a = 0.40$ fm at $\beta_1 = 6.8$.

\hspace*{12mm}

\begin{figure}[!h]
\begin{center}
\vspace*{-10mm}
\hspace*{11mm}
\epsfig{file=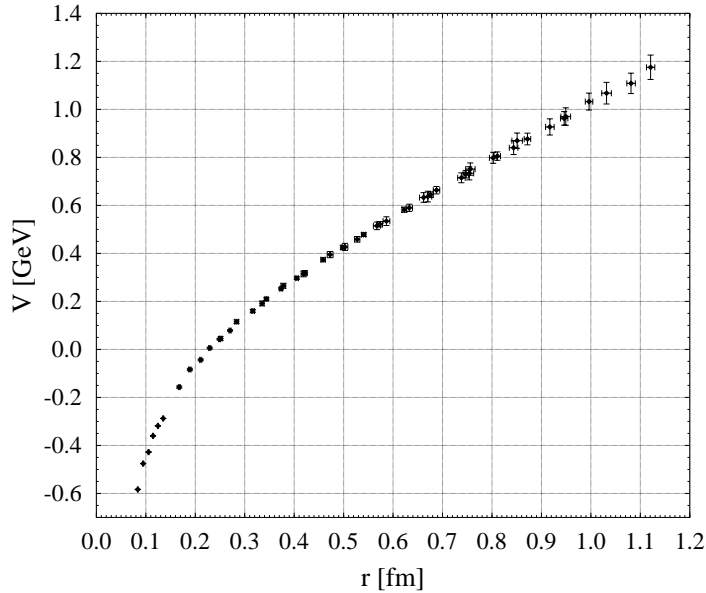,width=11.5cm,clip}
\caption{Superposition of the potentials for all
values of $\beta_1$. The irrelevant constant $C$ in Formula 
(\ref{contpot}) is set to zero.
\label{allpots}}
\end{center}
\end{figure}

\pagebreak

Finally in Fig.~\ref{allpots} we show a common plot of our results
for the static potential at all values of $\beta_1$ we analyzed. 
We set the irrelevant overall constant $C$ to zero for all $\beta_1$.
It is obvious that the data from different lattice 
spacings are in perfect 
agreement and the discretization errors are hardly noticeable for the 
L\"uscher-Weisz action. 
\\
\\
{\bf Acknowledgements: } We would like to thank Meinulf G\"ockeler,
Anna Hasenfratz, Francesco 
Knechtli, Paul Rakow and Andreas Sch\"afer for interesting 
discussions.

\end{document}